\documentclass[%
 reprint,
 amsmath,amssymb,
 aps,
]{revtex4-2}

\usepackage{graphicx}
\usepackage{dcolumn}
\usepackage{bm}

\begin{document}
	
	\preprint{APS/123-QED}
	
	\title{New perspective on thermodynamics of spacetime:\\ The emergence of unimodular gravity and the equivalence of entropies}
	
	\author{A. Alonso-Serrano}
	\email{ana.alonso.serrano@aei.mpg.de}
	\affiliation{Max-Planck-Institut  f\"ur  Gravitationsphysik  (Albert-Einstein-Institut),\\Am M\"{u}hlenberg 1, 14476 Potsdam, Germany}
	
	\author{M. Li\v{s}ka}
	\email{liska.mk@seznam.cz}
	\affiliation{Institute of Theoretical Physics, Faculty of Mathematics and Physics, Charles University,
	V Hole\v{s}ovi\v{c}k\'{a}ch 2, 180 00 Prague 8, Czech Republic}

	\begin{abstract}
		We present a novel derivation of Einstein equations from the balance between Clausius entropy crossing the boundary of a local causal diamond and entanglement entropy associated with its horizon. Comparing this derivation with the entanglement equilibrium approach developed by Jacobson, we are able to argue for the equivalence of  matter entanglement and Clausius entropy in the semiclassical regime. We also provide a direct comparison of both entropies for conformal matter, showing their equivalence without appealing to gravitational dynamics. Furthermore, we determine that gravitational dynamics implied by thermodynamics of spacetime, in fact, corresponds to unimodular gravity rather than general relativity.
	\end{abstract}

\maketitle

\section{Introduction}

The connection between thermodynamics and gravitational physics first appeared in the context of black hole thermodynamics~\cite{Bekenstein:1973,Bardeen:1973,Hawking:1975}. Since then, considerable attention has been devoted to understanding its nature and implications~\cite{Wald:2001}.

Significant progress in this area, although not free of controversy, was the derivation of Einstein equations from thermodynamics of (local virtual) Rindler horizons~\cite{Jacobson:1995ab}. It assumes that local Rindler horizons are in thermodynamic equilibrium, thus, matter entropy flux across the causal horizon is compensated by the change of quantum entanglement entropy associated with it (assumed to be proportional to the horizon's area). The matter entropy is defined via the equilibrium Clausius relation, $\text{d}S=\delta Q/T$, being $\delta Q$ the matter-energy flux across the causal horizon and $T$ the Unruh temperature. If the entanglement entropy associated with the horizon is identified with Bekenstein entropy, equilibrium conditions for the Rindler horizons constructed in every spacetime point imply Einstein equations.

Later, the notion of Clausius entropy was expanded to almost any null bifurcate surface~\cite{Baccetti:2013ica}, and a complete derivation of Einstein equations was carried out for stretched light cones~\cite{Svesko:2017}. Einstein equations were also obtained in a more rigorous way from the thermodynamic equilibrium of local causal diamonds, with matter entropy described in terms of entanglement entropy rather than by the Clausius relation~\cite{Jacobson:2015}. Since entanglement entropy is a purely quantum concept, the resulting equations of motion involve quantum expectation values rather than classical quantities. Let us further remark that the derivation for local Rindler horizons was also improved to work in the context of nonequilibrium thermodynamics~\cite{Eling:2006,Chirco:2010} and similar thermodynamic derivations were provided for the equations of motion of certain modified theories of gravity~\cite{Eling:2006,Chirco:2010,Padmanabhan:2010,Jacobson:2012,Bueno:2017,Svesko:2017,Svesko:2019}. Overall, thermodynamics of spacetime can be used not only to derive Einstein equations and equations of motion for several modified theories of gravity but can also provide insight into situations in which sources of gravity are quantum fields rather than classical matter. 

In the present paper, we provide an alternative derivation of Einstein equations from thermodynamics of local causal diamonds, using the construction of Clausius entropy flux across an arbitrary causal horizon presented in~\cite{Baccetti:2013ica}. Our derivation is based on well-known methods and definitions, and provides a new perspective on several unclear features of thermodynamics of spacetime. By using geodesic local causal diamonds (GLCDs) as the basic structure~\cite{Jacobson:2015} and describing matter entropy in terms of the semiclassical Clausius relation rather than quantum entanglement; we show the equivalence of Clausius and matter entanglement entropy. We also provide an argument for the equivalence of both entropies for a causal diamond filled with conformal matter, which is completely independent of gravitational dynamics. The relevance of this result lies in the fact that thermodynamic derivations of gravitational equations of motion usually consider entanglement entropy associated with a local causal horizon together with Clausius entropy flux across it~\cite{Jacobson:1995ab,Eling:2006,Chirco:2010,Jacobson:2012,Svesko:2017}. If Clausius entropy and matter entanglement entropy differ in the semiclassical regime, these derivations compare two conceptually different entropies. Here we partially resolve this issue, using the example of local causal diamonds to argue that such a comparison is indeed justified.

Furthermore, we find that the gravitational dynamics has much more in common with unimodular gravity (UG) than general relativity (GR). We argue that this is the case for any thermodynamic derivation of Einstein equations. Since Einstein equations can also be obtained from the equations of motion of UG (by assuming a divergence-free energy-momentum tensor), this in no way contradicts the previous thermodynamic results. Some aspects of thermodynamics derivation that can be associated with UG were already noted (without that interpretation) in the literature (see, e.g.~\cite{Padmanabhan:2010}). On writing this paper we also found that a broad connection between thermodynamics of spacetime and UG was already pointed out~\cite{Tiwari:2006}. Our approach shows the emergence of UG as a direct consequence of the derivation, as we automatically obtain a traceless right-hand side of the gravitational equations of motion (a characteristic feature of UG).

The paper is organised in the following way. First, in section~\ref{GLCD geometry}, the geometry of geodesic local causal diamonds is introduced, which is the setting of our thermodynamic considerations. In section~\ref{entropy}, we discuss all kinds of entropy involved in thermodynamics of spacetime and their relations. Especially, we present new arguments for the equivalence of Clausius and matter entanglement entropy. Section~\ref{dynamics} is concerned with obtaining gravitational dynamics from thermodynamics of spacetime. We first briefly review the entanglement equilibrium derivation in this setting, and then develop a novel derivation based on Clausius entropy flux. We also analyse the connection between thermodynamics of spacetime and UG. Finally, in section~\ref{discussion}, we discuss our results and outline some possibilities for future research.

Throughout the paper, we work in four spacetime dimensions and use metric signature $(-,+,+,+)$. Definitions of curvature-related quantities follow~\cite{MTW}. We use lower case Greek letters to denote abstract spacetime indices and lower case Latin letters for spatial indices with respect to a (local) Cartesian basis. Unless otherwise explicitly stated, we use Standard International units.

\section{Geodesic local causal diamonds}
\label{GLCD geometry}

Local causal diamonds are a natural setting for thermodynamic derivations of gravitational dynamics, being closed and easy to construct locally, without worrying about the horizon's continuation. Furthermore, a causal diamond is fully determined by its origin, a choice of a local time coordinate and a single length scale. In contrast, local Rindler wedges, usually considered in spacetime thermodynamics, require an arbitrary choice of a small pencil of horizon generators and, also, of an arbitrary interval of the null parameter along the generators. This makes Rindler wedges more cumbersome to work with and complicates generalisations of the formalism (see, e.g.~\cite{Svesko:2017,Svesko:2019} for further details).

For these reasons, our paper focuses on thermodynamic properties of GLCDs. We begin by introducing the construction of a GLCD and its basic properties. A more detailed discussion of causal diamonds can be found in, e.g.\mbox{~\cite{Gibbons:2007,Jacobson:2017,Wang:2019}.}

In any spacetime point, $P$, choose an arbitrary unit timelike vector, $n^{\mu}$, and construct Riemann normal coordinates (RNC), in which the metric nearby $P$ equals~\cite{Brewin:2009}
\begin{equation}
\label{RNC}
g_{\mu\nu}(x)=\eta_{\mu\nu}-\frac{1}{3}R_{\mu\alpha\nu\beta}\left(P\right)x^{\alpha}x^{\beta}+O\left(x^3\right).
\end{equation}

\begin{figure}[tbp]
	\centering
	\includegraphics[width=.45\textwidth,origin=c,trim={0.1cm 2.4cm 36.7cm 1.5cm},clip]{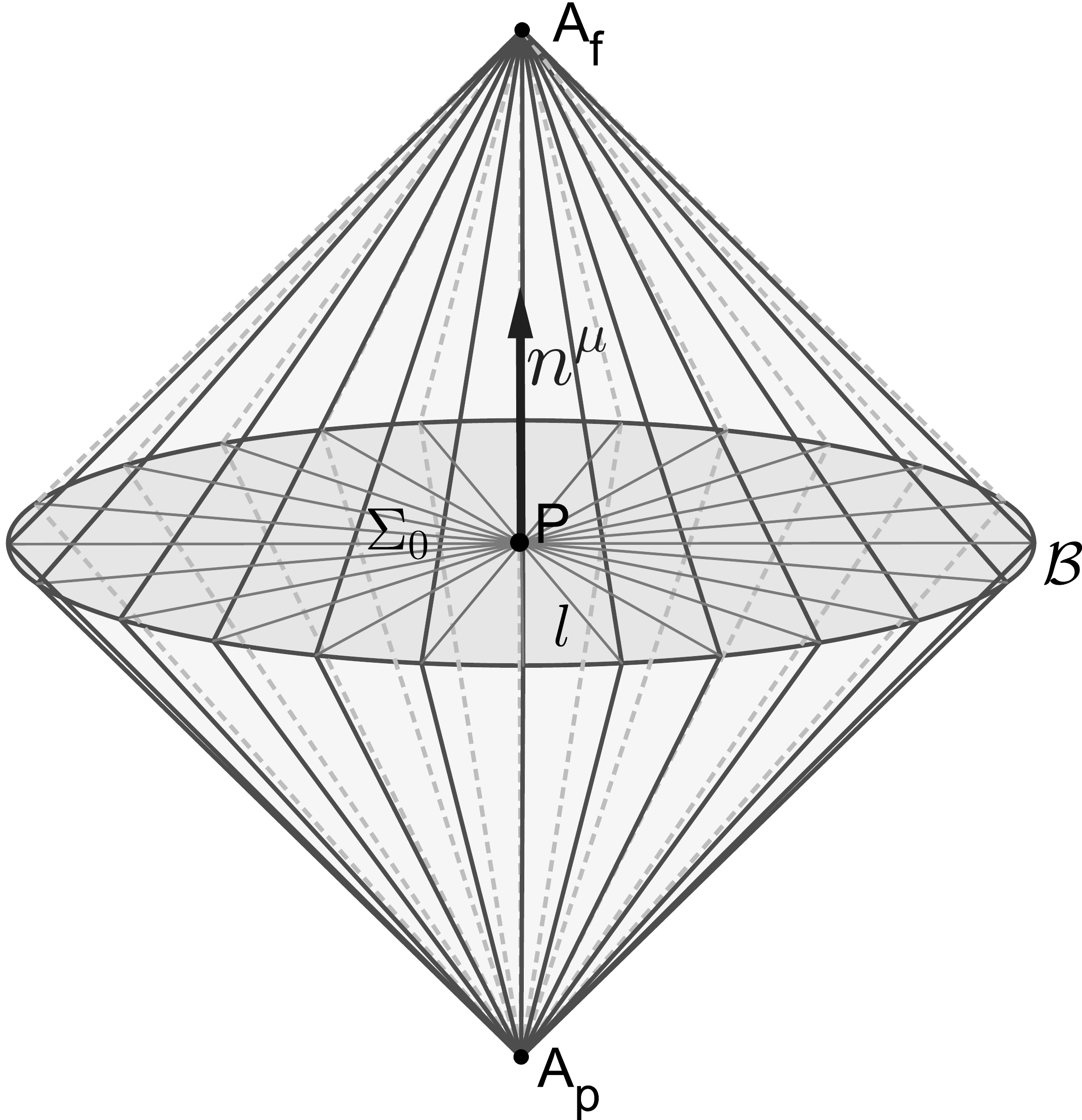}
	\caption{\label{diamond} A schematic picture of a GLCD with angular coordinate $\theta$ suppressed. Diamond's base $\Sigma_0$ is a spacelike geodesic ball, formed by geodesics of parameter length $l$ sent out from point $P$ (represented by the grey lines inside the base). Boundary $\mathcal{B}$ of $\Sigma_0$ is approximately a two-sphere. The ball is orthogonal to a timelike vector $n^\mu$. The tilted lines represent geodesic generators of the diamond's null boundary. The generators all start from past apex $A_p$ (corresponding to coordinate time $-l/c$) and again converge together in future apex $A_f$ (coordinate time $l/c$). Thus, the diamond's base is the spatial cross section of the future domain of dependence of $A_p$ at coordinate time $t=0$ and, likewise, the cross section of the past domain of dependence of $A_f$.}
\end{figure}

In every direction orthogonal to $n^{\mu}$, send out of $P$ geodesics of parameter length $l$, forming a three-dimensional geodesic ball, which we denote by $\Sigma_0$. The spacetime region causally determined by this ball is called a geodesic local causal diamond. For a clear visualisation, a sketch of GLCD is represented in figure~\ref{diamond}.

Assuming $l<<L_{curvature}$, where $L_{curvature}$ is the characteristic local curvature scale, i.e., inverse of the square root of the Riemann tensor's largest eigenvalue, the boundary, $\mathcal{B}$, of $\Sigma_0$ is approximately a two-sphere of area
\begin{equation}
\label{area1}
\mathcal{A}=4\pi l^2-\frac{2\pi}{9}l^4R^{ij}_{\;\:\:ij}\left(P\right)+O\left(l^5\right),
\end{equation}
where Latin indices denote the spatial components. From now on, we use that $x_{\mu}x^{\mu}\le l^2$ inside the ball to write the error terms only in powers of $l$. The extrinsic curvature of a geodesic ball vanishes up to $O\left(l\right)$~\cite{Jacobson:2017}. Thus, \mbox{$R^{ij}_{\;\:\:ij}={^{(3)}R}=2G_{00}+O\left(l\right)$~\cite{Jacobson:2015}}, where ${^{(3)}R}$ is the intrinsic scalar curvature, and the expression for the area results in
\begin{equation}\label{area}
\mathcal{A}=4\pi l^2-\frac{4\pi}{9}l^4G_{00}\left(P\right)+O\left(l^5\right),
\end{equation}
where $G_{00}=G_{\alpha\beta}n^{\alpha}n^{\beta}$.

In flat spacetime, a GLCD is endowed with a unique (up to a multiplicative constant) conformal Killing vector, generating a spherically symmetric conformal isometry that preserves the GLCD~\cite{Jacobson:2015}
\begin{equation}
\label{conformal Killing}
\xi=C\left(\left(l^2-t^2-r^2\right)\frac{\partial}{\partial t}-2rt\frac{\partial}{\partial r}\right).
\end{equation}
We can see that the null boundary of the GLCD is a conformal Killing horizon, as the vector becomes null there. In curved spacetime, the conformal symmetry is still satisfied up to $O\left(l^3\right)$.

\section{Entropy in thermodynamics of spacetime}
\label{entropy}

The key result of thermodynamics of spacetime is that gravitational dynamics is encoded in the equilibrium condition for maximal entropy: $\delta S=0$, valid for all virtual causal horizons (in our case null boundaries of GLCDs, constructed at every point of spacetime). This condition involves both entropy of quantum correlations across the horizon and that of matter-energy crossing it. Since, usually, the former is described in terms of quantum von Neumann entropy and the latter as thermodynamic Clausius entropy, it is not obvious that they can be combined to define a meaningful equilibrium condition. To prove this is the case, one would have to show that Clausius entropy flux equals that of von Neumann entropy of matter with sufficient precision. Then, the equilibrium condition can be restated as demanding maximal total von Neumann entropy of the system, $\delta S_{vN}=0$. In the following, we argue that Clausius and matter entanglement entropy crossing the boundary of a causal diamond are indeed equal to the leading order for the case of conformal fields. Let us remark that the comparison we perform is independent of gravitational dynamics. Therefore, we can use it to justify thermodynamic derivation of gravitational equations of motion without presenting a circular argument. We start in this section by introducing all the relevant entropies that will take part in the derivation, i.e., entanglement entropy associated with causal horizons, matter entanglement entropy and Clausius entropy.

First, we discuss an interpretation of the entropy associated with local causal horizons. It has been shown that to recover Einstein equations from thermodynamics, it must be equal to Bekenstein entropy~\cite{Jacobson:1995ab} (for modified theories of gravity, it needs to be the corresponding Wald entropy~\cite{Wald:1993,Padmanabhan:2010}), that was originally defined for black hole event horizons as~\cite{Bekenstein:1973,Bardeen:1973,Hawking:1975}
\begin{equation}
\label{BH_entropy}
S_{BH}=\frac{k_B\mathcal{A}}{4l_P^2},
\end{equation}
where $\mathcal{A}$ denotes the area of the black hole's event horizon, $k_B$ the Boltzmann constant and $l_P=\sqrt{G\hbar/c^3}$ the Planck length. Since the Bekenstein's original paper~\cite{Bekenstein:1973}, much attention has been devoted to finding a microscopic interpretation of black hole entropy~\cite{Wald:2001}. One of the proposals interprets it as a result of the quantum entanglement between the regions which are causally separated by the horizon. An observer on one side of the horizon cannot access information on the other side. Since vacuum fluctuations of quantum fields are correlated across the horizon, some information is inaccessible to the observer, leading to nonzero entanglement entropy. On the semiclassical level, entanglement entropy associated with a horizon is proportional to its area~\cite{Sorkin:1986,Srednicki:1993,Das:2008,Solodukhin:2011}
\begin{equation}
\label{entanglement}
S_e=\eta\mathcal{A},
\end{equation}
where $\eta$ can in principle depend on the position in spacetime~\cite{Chirco:2010}. However, to interpret Bekenstein entropy as entanglement entropy, $\eta$ must be assumed to have a universal value, $\eta=k_B/4l_P^2$.

This expression for entanglement entropy holds not only for black hole event horizons, but also for observer dependent causal horizons such as Rindler wedges and spherical horizons in flat spacetime~\cite{Solodukhin:2011}. It has been shown that it is possible to interpret the entire Bekenstein entropy as entanglement entropy~\cite{Solodukhin:2011, Jacobson:1994}; and it is usually done in studies concerned with thermodynamics of spacetime.

In order to describe thermodynamic equilibrium of a local causal horizon, one must further include the entropy of matter. To be consistent with the entanglement interpretation of the entropy associated with the horizon, one should describe matter entropy in terms of quantum entanglement. For the case of causal diamonds, this entanglement entropy can be explicitly evaluated for small perturbations from vacuum. Density operator $\rho$ corresponding to the vacuum state of quantum fields inside geodesic ball $\Sigma_0$ obeys $\rho=e^{-K/k_BT}/Z$, with $K$ being the so-called modular Hamiltonian, $Z$ the partition function and $T=\hbar lC/\pi k_Bc$ the Unruh temperature associated with the defined conformal Killing vector $\xi^{\mu}$. For conformal fields, a variation of the modular Hamiltonian corresponds to a variation of the local matter Hamiltonian~\cite{Jacobson:2019} which reads
\begin{equation}
\label{delta H}
\delta K=\delta H=\int_{\Sigma_0}\delta\langle T_{\mu\nu}\rangle\xi^{\mu}n^{\nu}\text{d}^2\Omega\text{d}r,
\end{equation}
where $n=\partial/\partial t$. This allows us to explicitly calculate the matter entanglement entropy variation in the following way
\begin{equation}
\label{conformal matter entropy}
\delta S_{m}=\frac{1}{T}\delta K=\frac{2\pi k_B}{\hbar c}\frac{4\pi l^4}{15}\delta\langle T_{00}\rangle+O\left(l^5\right).
\end{equation}
In the case of a nonconformal quantum field, $\delta K$ and $\delta H$ are no longer equal. However, when the field theory possesses a UV fixed point, their difference is only some spacetime scalar, $X$, and it holds~\cite{Jacobson:2015,Speranza:2016,Casini:2016}
\begin{equation}
\label{non-conformal matter entropy}
\delta S_{m}=\frac{2\pi k_B}{\hbar c}\frac{4\pi l^4}{15}\left(\delta\langle T_{00}\rangle+\delta X\right).
\end{equation}
where $\delta X$, in general, depends on $l$ and can even contain terms that become dominant for small $l$.

Within previous definitions, it is possible to use equilibrium condition for total entropy $\delta S_e+\delta S_m=0$ to obtain Einstein equations~\cite{Jacobson:2015} (we will review this derivation in subsection~\ref{classical_entanglement}). Then, one simply demands that the total entanglement entropy of the system is maximal for the equilibrium state. However, it is usual to assume~\cite{Jacobson:1995ab,Padmanabhan:2010,Chirco:2010,Svesko:2017}
\begin{equation}
\label{Clausius condition}
\delta S_e+\delta S_C=0,
\end{equation}
where $S_C$ is computed from the heat flux across the horizon by Clausius relation. Then, two conceptually different entropies are summed together and it is unclear whether this equation represents the correct condition on maximum of the total entropy. This question of the entropy equivalence has been pointed out, e.g. in~\cite{Baccetti:2013ica}, where a rigorous definition of Clausius entropy flux across any arbitrary causal horizon was provided. Using this definition for the case of causal diamonds, we are able to argue that $\delta S_C$ equals $\delta S_m$ to the leading order for conformal fields. Therefore, former equation indeed appears to be the correct thermodynamic equilibrium condition for local causal horizons. Since the technique by which the Clausius entropy flux is constructed is relevant both for its comparison with matter entanglement entropy and for the derivation of Einstein equations we introduce in subsection~\ref{classical_clausius}, we provide in the following a detailed overview.

\subsection{Clausius entropy flux}
\label{Clausius}

Consider a GLCD in curved spacetime as introduced in section~\ref{GLCD geometry}. Uniformly accelerating observers with a large acceleration, $a+O\left(1\right)$, moving inside GLCD have the following velocity and acceleration
\begin{eqnarray}
V^{\mu} &=c\left(\cosh\left(\frac{a\tau}{c^2}\right),-\sinh\left(\frac{a\tau}{c^2}\right)m^{i}\right)+O\left(l\right)\\
a^{\mu} &=a\left(\sinh\left(\frac{a\tau}{c^2}\right),-\cosh\left(\frac{a\tau}{c^2}\right)m^{i}\right)+O\left(1\right),
\end{eqnarray}
where $\tau$ is the observers' proper time and the correction terms come from the RNC expansion of the metric. The observers' worldlines intersect at coordinate times $t_{collision}=\pm\sqrt{l^2/c^2-c^2/a^2}$. Then, the approximately hyperbolic timelike sheet $\Sigma$ they sweep out has a unit normal
\begin{equation}
N^{\mu}=\left(-\sinh\left(\frac{a\tau}{c^2}\right),\cosh\left(\frac{a\tau}{c^2}\right)m^{i}\right)+O\left(l\right).
\end{equation}
The future pointing energy density flux vector crossing $\Sigma$ equals $-T^{\mu}_{\;\:\nu}V^{\nu}$. The total outward heat flux across a segment of the hypersurface reads
\begin{equation}
\label{heat}
\delta Q=-\int_{\Sigma}T_{\mu\nu}V^{\nu}N^{\mu}\text{d}^3\Sigma.
\end{equation}

To use the equilibrium Clausius relation $\text{d}S=\delta Q/T$, we further need a well-defined notion of temperature. 
Note that the Unruh effect for finite lifetime observers moving inside a causal diamond in flat spacetime can be rigorously defined~\cite{Martinetti:2003} (some of these results have been verified also using a different method~\cite{Arzano:2020}). It was found that for high accelerations the Unruh temperature obeys the standard formula, $T=\hbar a/2\pi k_Bc$. While acceleration of the observers is no longer exactly constant in curved spacetime, it has been shown that the Unruh effect will still hold adiabatically if the region where the acceleration is approximately constant is large compared to $c^2/a$~\cite{Barbado:2012}. Therefore, as long as we restrict ourselves to sufficiently large values of $a$, the use of the standard Unruh temperature in the Clausius relation is justified, and we obtain
\begin{equation}
\label{Clausius general}
S_{C}=\frac{\delta Q}{T}=-\frac{2\pi k_Bc}{\hbar a}\int_{\Sigma}T_{\mu\nu}V^{\nu}N^{\mu}\text{d}^3\Sigma+O\left(l^5\right).
\end{equation}
Note that this definition of Clausius entropy flux is clearly observer dependent.

In the limit of $a\to\infty$, the timelike sheet $\Sigma$ approaches the causal horizon of the GLCD. The coordinate time derivative of $S_C$ in this limit equals
\begin{equation}
\label{dS/dt null}
\frac{\text{d}S_{C}\left(t\right)}{\text{d}t}=\frac{2\pi k_Bc}{\hbar}t\int_{\mathcal{S}\left(t\right)}T_{\mu\nu}k_{\pm}^{\mu}k_{\pm}^{\nu}\text{d}^2\mathcal{A}+O\left(l^4\right),
\end{equation}
where $\mathcal{S}\left(t\right)$ is the spatial cross section of the horizon at time $t$ (approximately a two-sphere) and \mbox{$k_{\pm}^{\mu}=\left(1,-\text{sign}\left(t\right)m^{i}\right)+O\left(l\right)$}, are future pointing null normals to the causal horizon for $t>0$ and $t<0$, respectively, with $m^i=\left(\sin\theta\cos\phi,\sin\theta\sin\phi,\cos\theta\right)$ being the radial unit three-vector.  In flat spacetime, this equation holds exactly. Note that, on one hand, due to necessity to invoke the Unruh effect, the definition of Clausius entropy flux across a causal horizon requires the consideration of quantum mechanics. On the other hand, it is completely independent of gravitational dynamics.

Previous expression can be integrated to obtain the total Clausius entropy of the horizon at any given time (smaller than the intersection time, $t=l/c$). For the case of a GLCD, integration from the bifurcation two-sphere $\mathcal{B}$ at $t=0$ to the diamond's future apex $A_f$ at $t=l/c$ (see figure~\ref{diamond}) yields the total decrease of Clausius entropy
\begin{align}
\nonumber \Delta S_C=&S_C\left(\mathcal{B}\right)-S_C\left(A_f\right)\\
\nonumber =&-\frac{2\pi k_B c}{\hbar}\int_{0}^{\frac{l}{c}}t\int_{\mathcal{S}\left(t\right)}T_{\mu\nu}\left(x\left(t,\theta,\phi\right)\right)k_{+}^{\mu}k_{+}^{\nu}\text{d}^2\mathcal{A}\text{d}t \\
&+O\left(l^5\right),
\end{align}
where for the area element it holds \mbox{$\text{d}^2\mathcal{A}\left(t\right)=\left(l-ct\right)^2\text{d}^2\Omega+O\left(l^4\right)$}, with $\text{d}^2\Omega=\sin\theta\text{d}\theta\text{d}\phi$. To explicitly evaluate the integral, we expand the energy-momentum tensor around the origin of coordinates, $T_{\mu\nu}\left(x\left(t,\theta,\phi\right)\right)=T_{\mu\nu}\left(P\right)+O\left(l\right)$. Furthermore, we use $\int m^{i}\text{d}^2\Omega=0$, $\int m^{i}m^{j}\text{d}^2\Omega=4\pi\delta^{ij}/3$~\cite{Jacobson:2017}. This results into
\begin{equation}
\label{S_Clausius}
\Delta S_C=-\frac{2\pi^2 k_Bl^4}{3\hbar c}\left(T_{00}\left(P\right)+\frac{1}{3}T^{i}_{\;\:i}\left(P\right)\right)+O\left(l^5\right),
\end{equation}
which can be rewritten, using $T=T^{i}_{\;\:i}-T_{00}+O\left(l^2\right)$, in the form
\begin{equation}
\label{S_Clausius 2}
\Delta S_C=-\frac{8\pi^2 k_Bl^4}{9\hbar c}\left(T_{00}\left(P\right)+\frac{1}{4}T\left(P\right)\right)+O\left(l^5\right).
\end{equation}
One can expect the entropy of a single point, $A_f$, to be zero~\cite{Baccetti:2013ica}. Therefore, it holds $S_C\left(\mathcal{B}\right)=\Delta S_C$.

\subsection{Entropy equivalence}
\label{comparison}

Upon reviewing the definitions of matter entanglement entropy, $S_m$, and Clausius entropy, $S_C$, we are now ready to perform their comparison. We start by addressing the case of conformal fields and then turn our attention to the general situation. 

By comparing the computed values of both entropies, we immediately see that matter entanglement and Clausius entropy are proportional to each other (note that the energy-momentum tensor of a conformal field is traceless), but the numerical factors differ. However, the difference can be explained by the way both entropies are defined. On one hand, matter entanglement entropy, $S_m$, is computed inside the ball $\Sigma_0$, regardless of its development in time. On the other hand, $S_C$ represents the Clausius entropy of matter crossing the GLCD horizon. In fact, it can be easily shown that Clausius entropy is different for a light cone's spatial cross section of radius $l$ ($S_{C}=\left(8\pi^2k_Bl^4/3\hbar c\right)T_{00}$) and for the GLCD bifurcation surface of the same radius ($S_{C}=\left(8\pi^2k_Bl^4/9\hbar c\right)T_{00}$), although both surfaces are two-spheres of the same radius and $S_m$ would be the same for the inner regions of both. The reason for this is the preferred nature of the bifurcation surface $t=0$ in the calculation of $S_C$~\cite{Baccetti:2013ica} ($\text{d}S_{C}/\text{d}t=0$ on the bifurcation surface).

To properly compare $S_m$ and $S_C$, we calculate their coordinate time derivatives. As they quantify the momentary entropy flux across the horizon, the above-discussed differences between both entropies do not affect them. Before doing so, we must realise that $S_m$ depends on the conformal Killing vector $\xi^{\mu}$, whose definition changes with time, i.e., at time $t$, one should use conformal Killing vector $\xi'^{\mu}$ corresponding to GLCD of parameter length $l-t$, not the formula for $\xi^{\mu}$ evaluated at $t$. This also changes the definition of the Unruh temperature (which uses $\xi^{\mu}$). Therefore, the comparison of time derivatives can be only reliably done for small values of $t$ for which we can still use the original Killing vector $\xi^{\mu}$, i.e., by considering only the lowest order terms in $t$. On one side, for $S_C$, we get
\begin{equation}
\label{Clausius derivative}
\frac{\text{d}S_{C}}{\text{d}t}\left(t\right)=-\frac{32k_Bc}{3\hbar}t\left(l-ct\right)^2T_{00}\approx-\frac{32\pi^2k_Bc}{3\hbar}tl^2T_{00}.
\end{equation}
On the other side, for $S_m$, we obtain
\begin{align}
\label{matter entanglement derivative}
\nonumber \frac{\text{d}S_m}{\text{d}t}\approx&\frac{4\pi^2k_Bc}{\hbar l}\delta\langle T_{00}\rangle\frac{\text{d}}{\text{d}t}\int_{0}^{l-ct}r^2\left(l^2-c^2t^2-r^2\right)\text{d}r \\
\approx&-\frac{32\pi^2k_Bc}{3\hbar}tl^2\delta\langle T_{00}\rangle.
\end{align}
We see that the time derivatives are indeed equal for small $t$. The only difference is that $\text{d}S_C/\text{d}t$ depends on the classical energy-momentum tensor and $\text{d}S_m/\text{d}t$ on the quantum expectation value of its variation. Therefore, $S_m$ and $S_C$ appear to be equivalent in the leading order in $l$, and then comparing the changes in Clausius and horizon entanglement entropy is justified.

For nonconformal matter, the situation is more complicated. The originally conjectured general formula for matter entanglement entropy was equation~\eqref{non-conformal matter entropy} with $X=T/4$~\cite{Jacobson:2015}. This agrees with the definition of $S_C$ in the same sense as for conformal matter. However, our procedure in the above presented form cannot be applied directly to the more precise formulas for nonconformal $S_m$~\cite{Speranza:2016,Casini:2016}. Thus, the explicit equivalence of $S_m$ and $S_C$ in the most general cases remains an open issue.

\section{Thermodynamics of spacetime and gravitational dynamics}
\label{dynamics}

Once we have discussed the basic concepts involved in the formalism, we proceed with presenting two independent derivations of Einstein equations from thermodynamics of causal diamonds. The first one describes the entropy of both the horizon and matter in terms of quantum entanglement~\cite{Jacobson:2015}. In the second one, we follow the definition of Clausius entropy for bifurcating horizons~\cite{Baccetti:2013ica} (reviewed in subsection~\ref{Clausius}) and use it for a novel derivation of gravitational dynamics. By comparing both derivations, we are able to argue for the general equivalence of Clausius and matter entanglement entropy at the semiclassical level. Furthermore, we clearly illustrate that thermodynamically derived gravitational dynamics correspond to unimodular gravity.

\subsection{Derivation from the maximal vacuum entanglement hypothesis}
\label{classical_entanglement}

In order to compare it later with the Clausius entropy approach we develop in subsection~\ref{classical_clausius}, we first shortly review a derivation of Einstein equations from the variation of entanglement entropy around equilibrium for the case of four spacetime dimensions~\cite{Jacobson:2015} (a generalisation to an arbitrary dimension is straightforward). Since this derivation describes both the entropy associated with the horizon and that of matter inside it in terms of quantum entanglement, it raises no issues with potential comparing of two different kinds of entropy. We also point out previously unnoticed unimodular character of the resulting gravitational dynamics.

The starting point of the derivation is the maximal vacuum entanglement hypothesis (MVEH), which states: ``When the geometry and quantum fields are simultaneously varied from maximal symmetry, the entanglement entropy in a small geodesic ball is maximal at fixed volume~\cite{Jacobson:2015}.''

To be meaningful, MVEH requires a finite and universal prescription for the area density of vacuum entanglement entropy. If the density agrees with Bekenstein formula and only first order variations of the local vacuum state for quantum fields are considered, MVEH implies Einstein equations~\cite{Jacobson:2015}.

Consider the aforementioned GLCD construction carried out in a maximally symmetric spacetime (MSS). A small variation of quantum fields from their vacuum state leads to a change of matter entanglement entropy, $\delta S_m$, in the spatial geodesic ball $\Sigma_0$.

Note that, to define $\delta S_m$, one needs to invoke the Unruh effect. In order to have the quantum fields present in the GLCD in a thermal state corresponding to a well-defined Unruh temperature, we assume that their ground state can be locally approximated by Minkowski vacuum~\cite{Chirco:2010}. This amounts to assume Einstein equivalence principle (EEP), which states: ``Fundamental non-gravitational test physics is not affected, locally and at any point of spacetime, by the presence of a gravitational field~\cite{Casola:2015}.''

The variation of quantum fields induces a change of the spacetime metric and, therefore, of the GLCD horizon area and the entanglement entropy $S_e$ associated with it, expressed as $S_e=\eta\mathcal{A}$, as we have seen. In the following, we consider that $\eta$ is a universal constant, unaffected by the variation of quantum fields (otherwise, Einstein equations cannot be recovered~\cite{Chirco:2010}).

Intuitively, one would tend to choose a variation of the metric which leaves the size parameter $l$ unchanged. However, it was shown that such a variation gives incorrect numerical factor in Einstein equations~\cite{Jacobson:2015}. Instead, one must consider a variation of the metric which leaves fixed the volume of $\Sigma_0$. The variation of $S_e$ from MSS at the fixed volume of $\Sigma_0$ equals~\cite{Jacobson:2015}
\begin{align}
\label{dS_e}
\nonumber \delta S_e=&S_e-S^{MSS}_e=\eta\delta\mathcal{A} \\
=&-\eta\frac{4\pi}{15}l^4\left(G_{00}\left(P\right)+\lambda\left(P\right)g_{00}\left(P\right)\right)+O\left(l^5\right),
\end{align}
where we have used $G^{MSS}_{00}=-\lambda g_{00}$. The curvature scale $\lambda$ is approximately (that is, up to $O\left(l\right)$) constant inside the GLCD, but can depend both on the diamond's origin $P$ and on the length $l$.

According to MVEH, the total entropy variation vanishes to the first order, resulting in
\begin{align}
\label{entanglement_balance}
\nonumber &\frac{2\pi k_Bc}{\hbar}\frac{4\pi}{15}l^4\left(\delta\langle T_{00}\left(P\right)\rangle-\delta X\left(P\right)g_{00}\left(P\right)\right) \\
&-\eta\frac{4\pi}{15}l^4\left(G_{00}\left(P\right)+\lambda\left(P\right)g_{00}\left(P\right)\right)=0.
\end{align}
Due to EEP, this equation holds in every spacetime point $P$. Furthermore, the construction of GLCD in point $P$ holds for any arbitrary unit timelike vector $n^{\mu}$. Therefore, we can obtain equations for all the components of the Einstein tensor by carefully choosing ten different unit timelike vectors in any coordinate system. For example, in RNC we pick vectors $\left(1,0,0,0\right)$, $\left(\sqrt{2},1,0,0\right)$, $\left(\sqrt{3},\sqrt{2},0,0\right)$, $\left(\sqrt{3},1,1,0\right)$, $\left(2,\sqrt{2},1,0\right)$, $\left(\sqrt{6},2,1,0\right)$, $\left(2,1,1,1\right)$, $\left(\sqrt{5},\sqrt{2},1,1\right)$, $\left(\sqrt{6},\sqrt{2},\sqrt{2},1\right)$, $\left(\sqrt{7},\sqrt{3},\sqrt{2},1\right)$. Writing prior equation for each of the vectors leads to a system of ten independent equations which uniquely fix all components of the Einstein tensor. Thus, we can dispense with the contractions with $n^{\mu}$ and the dependence on $P$ and obtain a system of equations valid throughout the spacetime
\begin{equation}
\label{GR/UG}
\frac{2\pi k_Bc}{\hbar}\frac{4\pi}{15}l^4\left(\delta\langle T_{\mu\nu}\rangle-\delta Xg_{\mu\nu}\right)-\eta\frac{4\pi}{15}l^4\left(G_{\mu\nu}+\lambda g_{\mu\nu}\right)=0.
\end{equation}
In order to obtain Einstein equations, one must set \mbox{$\eta=k_B/4l_P^2$}. Thus, $S_e$ corresponds to Bekenstein entropy. Note that EEP still allows for the gravitational constant $G$ to be a function of the position in spacetime~\cite{Casola:2015}. Then, $\eta$ is no longer constant and Einstein equations cannot be recovered (one would have to consider the variation of $\eta$). To deal with this, we invoke the strong equivalence principle (SEP): ``All test fundamental physics (including gravitational physics) is not affected, locally, by the presence of a gravitational field~\cite{Casola:2015}.'' SEP then requires $G$ to be a universal constant~\cite{Casola:2015}.

To fix $\lambda$ one can take a trace of equations of motion, yielding
\begin{equation}
\lambda=\frac{1}{4}R+\frac{8\pi G}{c^4}\left(\frac{1}{4}\delta\langle T\rangle-\delta X\right)
\end{equation}
and providing traceless equations of motion
\begin{equation}
\label{unimodular}
R_{\mu\nu}-\frac{1}{4}Rg_{\mu\nu}=\frac{8\pi G}{c^4}\left(\delta\langle T_{\mu\nu}\rangle-\frac{1}{4}\delta\langle T\rangle g_{\mu\nu}\right).
\end{equation}
Since, rather than a classical energy-momentum tensor, $\delta\langle T_{\mu\nu}\rangle$ is the expectation value of the energy-momentum tensor of quantum fields present in the spacetime, the Einstein tensor should also be understood as a quantum expectation value~\cite{Jacobson:2015}. In other regards, the result corresponds to the equations of motion of unimodular gravity. If one further imposes the local energy-momentum conservation condition, $T_{\mu\;\:;\nu}^{\;\:\nu}=0$, one recovers Einstein equations
\begin{equation}
\label{Einstein}
G_{\mu\nu}+\Lambda g_{\mu\nu}=\frac{8\pi G}{c^4}\delta\langle T_{\mu\nu}\rangle,
\end{equation}
in which the cosmological term appears as an arbitrary constant of integration. Notably, scalar $\delta X$, which measures the nonconformality of quantum fields, does not affect gravitational dynamics. In other words, only the conformally invariant part of the energy-momentum tensor couples to gravity. All these properties are characteristic for unimodular theories, as we will discuss in more detail in subsection~\ref{UG/GR}.

\subsection{Derivation from Clausius entropy flux}
\label{classical_clausius}

In our preceding analysis of the relation between matter entanglement and Clausius entropy associated with a GLCD, it was not possible to argue for their general equivalence. However, we made no reference to gravitational dynamics. Our main aim here is to further our analysis by comparing gravitational dynamics derived from matter entanglement and Clausius entropy in the same geometric setup. We also confirm the general unimodular character of thermodynamically derived gravitational dynamics, noted in the previous subsection.

We argue that Einstein equations can be derived from equilibrium condition on entropy. Specifically, we compare the total decrease of Clausius entropy between times $t=0$ and $t=l/c$, already computed above, with the corresponding change in the entanglement entropy associated with the GLCD horizon. Recall that by doing so, we consider both EEP (we invoke the Unruh effect to define Clausius entropy) and the equivalence of Clausius and matter entanglement entropy in the leading order (so we do not compare conceptually different entropies). The latter condition was already justified for conformal matter.

The decrease in entanglement entropy $S_e$ from $t=0$ to $t=l/c$ associated with the horizon is directly proportional to the area of the bifurcation two-sphere $\mathcal{B}$, $\mathcal{A}$, and results in
\begin{equation}
\Delta S_e=-4\pi\eta l^2+\frac{4\pi\eta l^4}{9}G_{00}\left(P\right)+O\left(l^5\right).
\end{equation}
It is clear that $\Delta S_e$ is nonzero even in flat spacetime in which $S_C$ is identically zero ($S_C$ arises only due to the presence of matter). Therefore, some ``equilibrium state contribution'' must be subtracted before the comparison of $\Delta S_e$ and $\Delta S_C$ is made. A similar problem was encountered in the derivation of gravitational dynamics from thermodynamics of stretched light cones (a light cone expands even in flat spacetime, and the increase in entropy due to this had to be subtracted)~\cite{Svesko:2017}. In previous subsection, we showed that the appropriate equilibrium state of GLCD, that allowed the derivation of Einstein equations, was a maximally symmetric spacetime, rather than a flat one; if the present quantum fields were nonconformal. As we made no assumptions about conformality, we take the most general approach and subtract a contribution corresponding to MSS,
\begin{equation}
\Delta S_{e\,MSS}=-4\pi\eta l^2-\frac{4\pi\eta l^4}{9}\lambda\left(P\right)g_{00}\left(P\right)+O\left(l^5\right),
\end{equation}
where $\lambda$ is again approximately constant inside the GLCD but can depend both on $P$ \mbox{and $l$.}

In total, we have the following thermodynamic equilibrium condition:
\begin{equation}
\Delta S_{e}-\Delta S_{e\,MSS}+\Delta S_{C}=0, \label{clausius_balance_1}\\
\end{equation}
which leads to
\begin{align}
\nonumber &G_{00}\left(P\right)+\lambda\left(P\right)g_{00}\left(P\right)-\frac{2\pi k_B}{\hbar c\eta} \\
&\times\left(T_{00}\left(P\right)-\frac{1}{4}T\left(P\right)g_{00}\left(P\right)\right)=0.
\end{align}
Note that EEP ensures its validity in every spacetime point $P$. Furthermore, it holds for any timelike vector $n^{\mu}$. Therefore, by the same argument as in the previous subsection, we obtain a system of tensorial equations
\begin{eqnarray}
\label{Einstein/unimodular}
G_{\mu\nu}+\lambda g_{\mu\nu} &=\frac{2\pi k_B}{\hbar c\eta}\left(T_{\mu\nu}-\frac{1}{4}Tg_{\mu\nu}\right).
\end{eqnarray}
Notice that only traceless, conformal part of the energy-momentum tensor appears on the right-hand side. Upon fixing $\lambda$ by taking a trace of these equations, we again find traceless equations of motion of unimodular gravity
\begin{equation}
\label{unimodular 1}
R_{\mu\nu}-\frac{1}{4}Rg_{\mu\nu}=\frac{8\pi G}{c^4}\left(T_{\mu\nu}-\frac{1}{4}Tg_{\mu\nu}\right).
\end{equation}
As before, $\eta$ is set to the Bekenstein value, $\eta=k_B/4l_P^2$ (note that we must again consider SEP, so that $G$ is a universal constant). Finally, the local energy-momentum conservation allows us to recover Einstein equations with $\Lambda$ appearing as an integration constant
\begin{equation}
G_{\mu\nu}+\Lambda g_{\mu\nu}=\frac{8\pi G}{c^4}T_{\mu\nu}.
\end{equation}

Note that both derivations yield completely equivalent gravitational dynamics in the same setting, despite using different descriptions of matter entropy. Furthermore, in both cases we consider the same maximally symmetric equilibrium state with locally constant curvature scale $\lambda$. For MVEH we have (if we demand local energy-momentum conservation) $\lambda=-\left(8\pi G/c^4\right)\delta X+\Lambda$ and for the Clausius entropy case $\lambda=-\left(2\pi G/c^4\right)T+\Lambda$. Note that in both cases $\lambda=0$ for conformal matter. Taking $\delta X=-T/4$ would make $\lambda$ the same even for nonconformal matter, but this is not generally the case~\cite{Speranza:2016,Casini:2016}. However, $\lambda$ in both cases contains the nonconformal contribution to matter entropy. We can conclude that only the conformally invariant sector of matter entropy couples to gravity, and this sector is the same both for Clausius and entanglement entropy. Therefore, both entropy descriptions lead to completely equivalent gravitational dynamics in the semiclassical limit, so they can be used interchangeably in thermodynamics of spacetime. Notably, the gravitational dynamics we find is in both cases unimodular, as we clearly show in the following.

\subsection{Thermodynamics of spacetime and unimodular gravity}
\label{UG/GR}

Traceless equations found in the previous subsection are precisely the equations of motion of unimodular gravity. The basic idea of UG is restricting the full diffeomorphism invariance of GR only to the diffeomorphisms keeping the determinant of the metric fixed, originally to $g=-1$. More generally, one can demand $g=-\epsilon_0^2$, with $\epsilon_0$ being an arbitrary real number. The simplest unimodular action reads
\begin{equation}
\label{unimodular action}
S=\int_{\Omega}\left(\frac{c^4}{16\pi G}R\left(\hat{g}_{\mu\nu}\right)+\mathcal{L}_{matter}\right)\epsilon_0\text{d}^4x,
\end{equation}
where $\Omega$ is a spacetime manifold, $\mathcal{L}_{matter}$ is the matter Lagrangian density, and $\hat{g}_{\mu\nu}$ denotes metrics satisfying the fixed determinant condition, $\hat{g}=-\epsilon_0^2$. Using integration by parts, one can rewrite the first term in this action in the following form~\cite{Barcelo:2018}:
\begin{align}
\label{unimodular action pp}
\nonumber S=&\int_{\Omega}\bigg(\frac{c^4}{16\pi G}\hat{g}^{\alpha\beta}_{\quad;\mu}\hat{g}^{\kappa\lambda}_{\quad;\nu}\left(2\hat{g}_{\alpha\kappa}\delta^{\mu}_{\lambda}\delta^{\nu}_{\beta}-\hat{g}^{\mu\nu}\hat{g}_{\alpha\kappa}\hat{g}_{\beta\lambda}\right)\\
&+\mathcal{L}_{matter}\bigg)\epsilon_0\text{d}^4x,
\end{align}
where the semicolon denotes covariant derivatives defined with respect to a background Minkowski metric, $\eta_{\mu\nu}$. The action can also be restated in terms of metric $g_{\mu\nu}=\left(-g/\epsilon_0^2\right)^{1/4}\hat{g}_{\mu\nu}$, whose determinant is arbitrary. This version of unimodular action is invariant under conformal transformations of the form
\begin{equation}
g'_{\mu\nu}=\Omega^2g_{\mu\nu},
\end{equation}
and under transverse diffeomorphism transformations, that are infinitesimally expressed
\begin{subequations}
\begin{eqnarray}
\hat{g}'_{\mu\nu}&=&\hat{g}_{\mu\nu}+2\xi_{(\mu;\nu)},  \\
\xi^{\mu}_{\;\:;\mu}&=&0.
\end{eqnarray}
\end{subequations}
The theory corresponding to this action is also known as Weyl transverse gravity~\cite{Barcelo:2018}.

The full diffeomorphism invariance of GR is lost due to the presence of a nondynamical proper volume element, $\text{d}\mathcal{V}=\epsilon_0\text{d}^4x$. Therefore, the action does not imply a divergence-free energy-momentum tensor that must be assumed as an additional requirement (although it is also possible to work without that assumption~\cite{Josset:2017}). Then, it is easy to rewrite the traceless equations of motion of UG as Einstein equations. 

The presence of a cosmological constant term in the Lagrangian breaks the invariance under conformal transformation~\cite{Valea:2016}. Moreover, any such term does not enter into classical equations of motion (although it becomes relevant on the quantum level~\cite{Fiol:2010}). Instead, the cosmological constant present in Einstein equations arises as an arbitrary integration constant from the local energy-momentum conservation condition, $T_{\mu\;\:;\nu}^{\;\:\nu}=0$. Apart from the different status of the cosmological constant, classical gravitational dynamics implied by UG and GR is equivalent. However, on the quantum level, diverse proposals regarding the equivalence of both theories appear in the literature (see, e.g.~\cite{Fiol:2010,Padilla:2014,Bufalo:2015}). Aside from the above cited works, more detailed treatment of UG can be found, e.g. in~\cite{Barcelo:2014}.

Now, one can see that both methods of obtaining gravitational dynamics from thermodynamics point to UG rather than GR. First, the cosmological term appears in both approaches due to integration of the local energy-momentum conservation condition. Therefore, the value of cosmological constant is arbitrary and equations of motion are invariant under simultaneous shifting of the energy-momentum tensor by $Cg_{\mu\nu}$ and the cosmological constant by $C$, being $C$ an arbitrary constant. This precisely corresponds to unimodular gravity. Second, conformal invariance breaking contributions to matter entropy, i.e., scalar $\delta X$ for MVEH derivation and the trace of the energy-momentum tensor for the Clausius flux approach, do not influence equations of motion. Therefore, only the conformally invariant (traceless) part of the energy-momentum tensor couples to gravity, allowing for invariance of equations of motion under conformal transformation. This is again consistent with the conformal invariance of UG action discussed above (for details of conformal invariance of UG in the presence of matter, see e.g.~\cite{Alvarez:2013}). Furthermore, unimodular condition $g=-1$ is locally obeyed  up to $O\left(l^2\right)$ by the RNC metric expansion considered in both derivations.

The same reasoning applies even to the original Jacobson's derivation of Einstein equations from thermodynamics of local Rindler wedges~\cite{Jacobson:1995ab} (and, indeed, to all thermodynamic derivations known to the authors). Without assuming $T_{\mu\;\:;\nu}^{\;\:\nu}=0$, it implies
\begin{equation}
\label{original Jacobson}
R_{\mu\nu}+\Phi g_{\mu\nu}=\frac{2\pi k_B}{\hbar c\eta}T_{\mu\nu},
\end{equation}
with $\Phi$ being an undetermined scalar function, which is invariant under adding $Cg_{\mu\nu}$ to $T_{\mu\nu}$ and simultaneously shifting $\Phi$ by $C$. Furthermore, $\Phi$ can simply be fixed by taking the trace of the equations, leading again to traceless equations of motion.

The reasons for unimodular nature of thermodynamically derived gravitational dynamics can be partially understood by considering properties of entropy. One way to see this is by noting that for the study of entropy is relevant only its difference between two states of the system and not its total value. This results into vacuum contribution to matter entanglement entropy being an arbitrary universal constant and, thus, its value does not affect the conditions for thermodynamic equilibrium. Then, in gravitational dynamics obtained from thermodynamics, vacuum energy naturally would not couple to gravity, leading to the behaviour of the cosmological constant characteristic for UG. This issue will be addressed in detail in a future work.

The similarities between thermodynamically derived gravitational equations of motion and UG were previously noted in~\cite{Tiwari:2006} by conjecturing a qualitative relation for the case of local Rindler wedges.  In~\cite{Padmanabhan:2010} the invariance of equations of motion under the shift of the energy-momentum tensor by $Cg_{\mu\nu}$ was pointed out (and implications for the nature of the cosmological constant were discussed), but the connection with UG was not explored. So far, we have only discussed classical equations of motion, which are equivalent in both UG and GR. To reliably decide toward which theory the thermodynamic methods point, one would have to analyse quantum corrections to equations of motion, that break the equivalence of both theories. We carry out such an analysis in a paper under preparation, in which quantum modified dynamics are obtained by considering logarithmic corrections to entanglement entropy~\cite{preparation}.

\section{Discussion}
\label{discussion}

Building on the construction of Clausius entropy for bifurcating horizons~\cite{Baccetti:2013ica}, we present a novel derivation of Einstein equations from thermodynamics of local causal diamonds. By showing that this derivation is fully equivalent to the one based on MVEH~\cite{Jacobson:2015}, we showed that Clausius entropy~\cite{Baccetti:2013ica} and matter entanglement entropy can be used interchangeably in thermodynamics of spacetime in the semiclassical regime. For the case of conformal matter, we even explicitly prove the equivalence without appealing to gravitational physics in any way. This retrospectively justifies the use of Clausius entropy instead of matter entanglement entropy in the Jacobson's original paper on thermodynamics of spacetime~\cite{Jacobson:1995ab} and many subsequent works (e.g.,~\cite{Chirco:2010,Padmanabhan:2010,Svesko:2017}).

Furthermore, our method provides a connection between the maximal vacuum entanglement hypothesis approach~\cite{Jacobson:2015} and the original Jacobson's thermodynamic derivation~\cite{Jacobson:1995ab}, as it reproduces the results obtained from MVEH in the same setting but using Clausius rather than matter entanglement entropy. In other words, we have employed the construction of Clausius entropy from~\cite{Jacobson:1995ab} (although in a more refined formulation of~\cite{Baccetti:2013ica}) in the geometric setting of~\cite{Jacobson:2015}, and we have found a result completely consistent with both approaches.

We also clearly show the unimodular nature of thermodynamically derived gravitational dynamics. This is evident from the fact that gravity couples only to the conformal part of the energy-momentum tensor and that vacuum energy does not gravitate. It appears that these features are encoded in the properties of entropy, but the precise nature of this encoding deserves further study. Moreover, we need to better understand how does the loss of full diffeomorphism invariance and the presence of background structures in UG relate to the thermodynamic description. Furthermore, it appears that thermodynamics of spacetime allows one to consider certain energy-momentum tensors that are not divergence free. It might be worthwhile to study what this in turn implies for the thermodynamic description. Let us also remark that the different status of the cosmological constant in this approach might deserve further attention.

In a future work, we will address the issue of showing the entropy equivalence in a strict way for nonconformal matter and for generic local causal horizons. This would allow us to understand in which cases it is possible to derive Einstein equations from thermodynamic equilibrium conditions. Furthermore, we should not forget that both our GLCD derivation and the entropy equivalence question could be treated in case of nonequilibrium thermodynamics of spacetime (and possibly of modified theories of gravity).

The scope of the present analysis is limited to semiclassical thermodynamics of spacetime. Therefore, due to the equivalence of the classical dynamics implied by UG and GR, we are unable to conclusively prove the unimodular nature of the thermodynamically derived gravitational dynamics. However, this should be possible when quantum corrections to the equations of motion are considered. In that case, we also explore whether the corrections implied by quantum gravity effects break the entropy equivalence~\cite{preparation}.

\acknowledgments 

The authors want to acknowledge Matt Visser and Luis J. Garay for reading a first draft of the paper and giving their feedback. Also, the authors acknowledge discussions with \mbox{Jos\'{e} M.M. Senovilla} and with the participants of the \mbox{CARRAMPLAS} workshop.

A. A.-S. is funded by the Alexander von 	Humboldt  Foundation. The work of A. A.-S. is also partially supported by the Project. No. MINECO FIS2017-86497-C2-2-P from Spain. This paper is based upon work from COST Action CA15117 ``Cosmology and Astrophysics Network for Theoretical Advances and Training Actions (CANTATA),'' supported by COST (European Cooperation in Science and Technology).

\end{document}